\documentclass[11pt]{article}

\usepackage{graphicx}
\usepackage{amssymb}
\usepackage{amsmath}
\usepackage[symbol]{footmisc} 
\usepackage{enumitem}
\usepackage{color}
\usepackage{titlesec}
\usepackage{ulem}
\usepackage{authblk}  
\usepackage{caption}
\usepackage{dcolumn}
\usepackage{bm}
\usepackage{wasysym}
\usepackage{appendix}
\usepackage[mathlines]  {lineno}

 \titlespacing*{\subsection}{0pt}{10pt}{-5pt}
\titlespacing*{\section}{0pt}{20pt}{5pt}
\newcommand{\be}{\begin{equation}}
\newcommand{\ee}{\end{equation}}
\newcommand{\benn}{\begin{equation*}}
\newcommand{\eenn}{\end{equation*}}
\newcommand{\ba}{\begin{eqnarray}}
\newcommand{\ea}{\end{eqnarray}}

\newcommand{\larghez}{\columnwidth}
\def\nn{\nonumber}

\begin{document}
   \title{Extra force in charged  resonant capacitors: \\a new macroscopic effect of vacuum fluctuations ?}
\author{Yury Minenkov}
\affil{\small{Istituto Nazionale di Fisica Nucleare (INFN), Sezione  Tor Vergata, 
Roma, Italy \\
yury.minenkov@roma2.infn.it }}

\author{Massimo Bassan}
\affil{\small{Dipartimento di Fisica,  Universit\`a di Tor Vergata and \\
Istituto Nazionale di Fisica Nucleare (INFN), Sezione  Tor Vergata, 
Roma, Italy\\
bassan@roma2.infn.it}}
\date{}
\maketitle
%

\begin{abstract}
We report on measurements of the mechanical vibrations of the fundamental mode of resonant capacitive transducers, consisting of parallel plates at different gaps (10 - 26 $\mu$m). Resonant frequency data were taken vs the electric field at constant charge, at various temperatures and different configurations. All measurements exhibited  unexpectedly large tuning, caused by an additional, unpredicted attractive force. Such  force 
has a divergent behaviour characterized by a critical electrical field and start to diverge long before the critical point. 
Many known physical mechanisms have been considered to explain this force, but all of them had to be discarded. 
Thus, we turn to non-conventional physical mechanisms and suggest that
vacuum fluctuations inside the gap 
could be a possible cause of this anomalous force.
\end{abstract}

\section{Introduction - What is the issue ?}
\label{sec:intro}
{\it We have evidence of a  force between the plates of a charged capacitor that we cannot account for with normal electromechanical arguments.} \\

We are considering frequency measurements on capacitive resonant vibration transducers, devices that were used, in a recent past, on cryogenic resonant g.w. detectors \cite{Expl,Naut}. These are capacitors formed by parallel disks  with  large surface ($S= 137 ~cm^2$) and small gap ($d \sim $10 to 26 $\mu m$), where one  massive ($m \simeq$ 0.36 to 0.64 kg) plate  can vibrate, in its fundamental mechanical mode of oscillation, in a direction normal to the plate surface, say $\hat z$, at an acoustic frequency $f_0 \sim 1$ kHz. \\
There are well known forces acting in the transducer:  the mechanical restoring force $F_{m}  $ and the electrostatic attraction between the two plates $F_{el}$. The latter has both a d.c. component and a time dependent one,  at the vibration frequency, as described below. Our tuning measurements lead us to believe that an additional, unexpected force is acting on the charged resonator.\\
Focusing  on the oscillating force, and neglecting friction for the moment, we write the equation of motion for the vibrating plate, that defines its resonant frequency:
\begin{equation}
m  \ddot z =  F_{m} + F_{el} + F_{un}
\label{eq:eqmotion}
\end{equation}
where the third term is the unexplained effect we discuss here.  
The tests were performed on many measuring sets, using  four different resonators, in six different mechanical set-ups and at  three, very different, temperatures,
as detailed in table \ref{tab1}.  Frequency measurements can achieve great accuracy:  in our case the resonant frequency is measured to 10 ppm or better, while the deviations we record are up to 700 \%.
In this note we recap the origin of the "normal"  tuning mechanism and we try to investigate the unexpected behaviour exhibited in our tests. 

\section{The tuning of a charged resonator}
\subsection{The ``normal" tuning}
\label{sec:normaltuning}
When an electromagnetic field is stored in the gap between two plates (one of them vibrating at $f_0$) of a capacitor, the energy associated with the field  produces an e.m. force $F_{el}$.  
If the force exhibits a gradient, we have an  ``electric stiffness"   $ K_{el}  =  \partial F_{el} / \partial z$: this varies the resonant frequency of the vibrating plate.
While we defer the description of the apparatus to the next section, we start with a few considerations useful to discuss the forces acting on the charged resonator: \\
-  In the fundamental mode of oscillation we are concerned with,  one electrode plate, the resonator, moves as a solid object of mass $m$ on the spring constant $ K_{m}$ provided by the supporting cantilevers (see fig.\ref{fig:foto}). 
The plate moves parallel to its rest position, along the direction $\hat z$ of its symmetry axis.  Internal modes of the plate occur at much higher frequency than $ \omega_0 =2\pi f_0  = \sqrt{K_m/m}$, so that the lumped oscillator approximation is fully appropriate.
  The elastic restoring force acting on the resonator is  $\vec F_{m} = -  K_{m} \vec z$, where, again,  $ K_{m}$ is the stiffness provided by the six cantilever arms.
Experimental determination of the elastic constant $ K_{m}$ is detailed in Appendix A.\\
- 
The capacitor $C = \varepsilon_0 S /d  \simeq 5 - 12~nF$  is charged, via two large  bias resistors (see fig.\ref{fig:schema})  by a voltage generator that is then disconnected.  An attractive force develops between the plates, changing their distance from $d_0$ (gap of the uncharged capacitor) to a new  equilibrium value $d_{eq}$ (see below). \\
- Even if we leave the generator connected (with $R = 2~10^7 ~\Omega$), the charge changes with time constant $\tau = RC  \simeq 0.1 s \gg 1/f_0$, so that  we can always consider 
 the system to operate  \uline{ in constant charge conditions}.  \\
- An unavoidable stray capacitance $C_s$, due to spacers and cables, appears in parallel with the "active" capacitor.

When the plate vibrates, $d(t) = d_{eq} + z(t)$, the capacitance changes and charge flows between active and stray capacitors.
We  write  the potential energy  as the sum of an electrical term and a mechanical one:

\begin{equation}
\mathcal E_p(z) = \frac12 \frac {Q^2}{ ( \frac{\varepsilon_0 S}{d_{eq} +z} + C_s)} +  \frac12 K_m (d_{eq} -d_0 + z)^2
\label{eq:energy}
\end{equation}

We can expand the energy in power series ( $z/d_{eq} \ll 1$) around $z = 0$:
\be
\mathcal E_p =  e_0 + e_1 z + \frac 12 e_2 z^2 + \frac 16 e_3 z^3 + \dots 
\label{eq:expand}
\ee
The force acting on the resonator is then:
\be
F_z= - \frac{d \mathcal E_p}{dz} = -e_1 -  e_2 z -  \frac12 e_3 z^2 ~+\dots
\label{eq:force}
\ee

We express the coefficients $e_i$ in terms of the applied (and measurable) voltage $V = Q/(C+C_s)$.
\begin{align}
F_z = &  - \bigg[\frac{C   V^2 }{2 d_{eq} }   +  K_m (d_{eq}-d_0) \bigg] ~~~+  \nn \\   
- &~ \bigg[ K_m  - \frac{V^2}{d_{eq}^2} \frac{C C_s}{(C+C_s)}   \bigg] \cdot z ~+ \nn \\
 + & ~\bigg[\frac32 ~~\frac{V^2}{d^3_{eq}} ~~\frac {C C_s^2}{(C + C_s)^2}  \bigg]~ \cdot z^2 + \dots
 \label{eq:force2}
\end{align}

 The  term in the first line  is the static force: balancing the electrostatic attraction  between the charged plates and the elastic push of the cantilevers determines the new equilibrium distance $d_{eq} $ between the plates, as we shall discuss in sect.\ref{sec:shrink}. \\
The second term is the elastic restoring force and it exhibits,
beside the mechanical stiffness $K_m$, an additional  term, depending on the electric field $E=V/d_{eq}$,  that we call ``electric stiffness"  \footnote{ The electric stiffness, defined as $ K_{el} \equiv  \partial^2 \mathcal E_{el} / \partial z^2 $ should in principle be negative. Here and in what follows we'll take, for ease of reading, its absolute value, so that its minus sign will be explicitly displayed.  Same, in the following, for the extra stiffness $K_{un}$. }
\be K_{el} \equiv   {E^2}~ \frac{C C_s}{(C+C_s)} 
\label{eq:Kel}
\ee
 This is a typical feature of all electromechanical transducers: alternate derivations are given in sect.\ref
 {sec:emIssues} and in Appendix C.
 In electrostatic devices, the  additional stiffness gives a negative contribution,
as explicitly shown  in eq.\ref{eq:force2},  and lowers the resonant frequency.
\begin{equation}
\omega^2(V) =  \frac{K_m - K_{el}(V)}{m} = \omega_0^2 - \frac{K_{el} (V)} {m}
\label{eq:omegatuning}
\end{equation}
We rewrite the tuning relation in terms of measurable quantities:
\begin{equation}
f^2_c(V) =  f_0^2 -  \frac1{4 \pi^2}~\frac{V^2}{d_{eq}^2m}~ \frac{C_s}{(1+C_s/C)}
\label{eq:tune}
\end{equation}
where $f_c$ refers to the expected (computed)  resonant frequency of the charged resonator. \\
Note that the second derivative $e_2$  is non-zero  due to the presence of $C_s$: else, the electric field would be uniform and constant in the gap and there would be no force gradient.  As the plate vibrates, modulating the active capacitance C,  electric charge flows between C and $C_s$,  the E field in the gap changes, giving rise to a time dependent, restoring force.\\
The expected tuning has a quadratic dependence on $V$ and is shown by the blue curves in fig.\ref{fig:tuning}.
In the following, we shall discuss either the tuning 
$(f^2_0 - f_c^2)$ or the stiffness relation $ K_{el}/ K_m$: the two quantities refer to the same physical effect, and are related by eq.\ref{eq:omegatuning}, 
that we can rewrite as 
\be
K_{el}/ K_m = \Delta f^2_c /f_0^2= 1- f^2_c /f_0^2
\ee
Finally, the   force term  in the third line of eq.\ref{eq:force2} 
 will only be considered in sect.\ref{sec:nonlinear}, when we deal with possible non-linearities. \\

\subsection{ The extra tuning, unaccounted for}
 We have seen that the resonant frequency of the oscillator is expected to vary when the capacitor is charged, and 
 we can predict the tuning curve based on eq.\ref{eq:tune}.
 \begin{figure}[h]
\begin{center}
\includegraphics[width=\larghez]{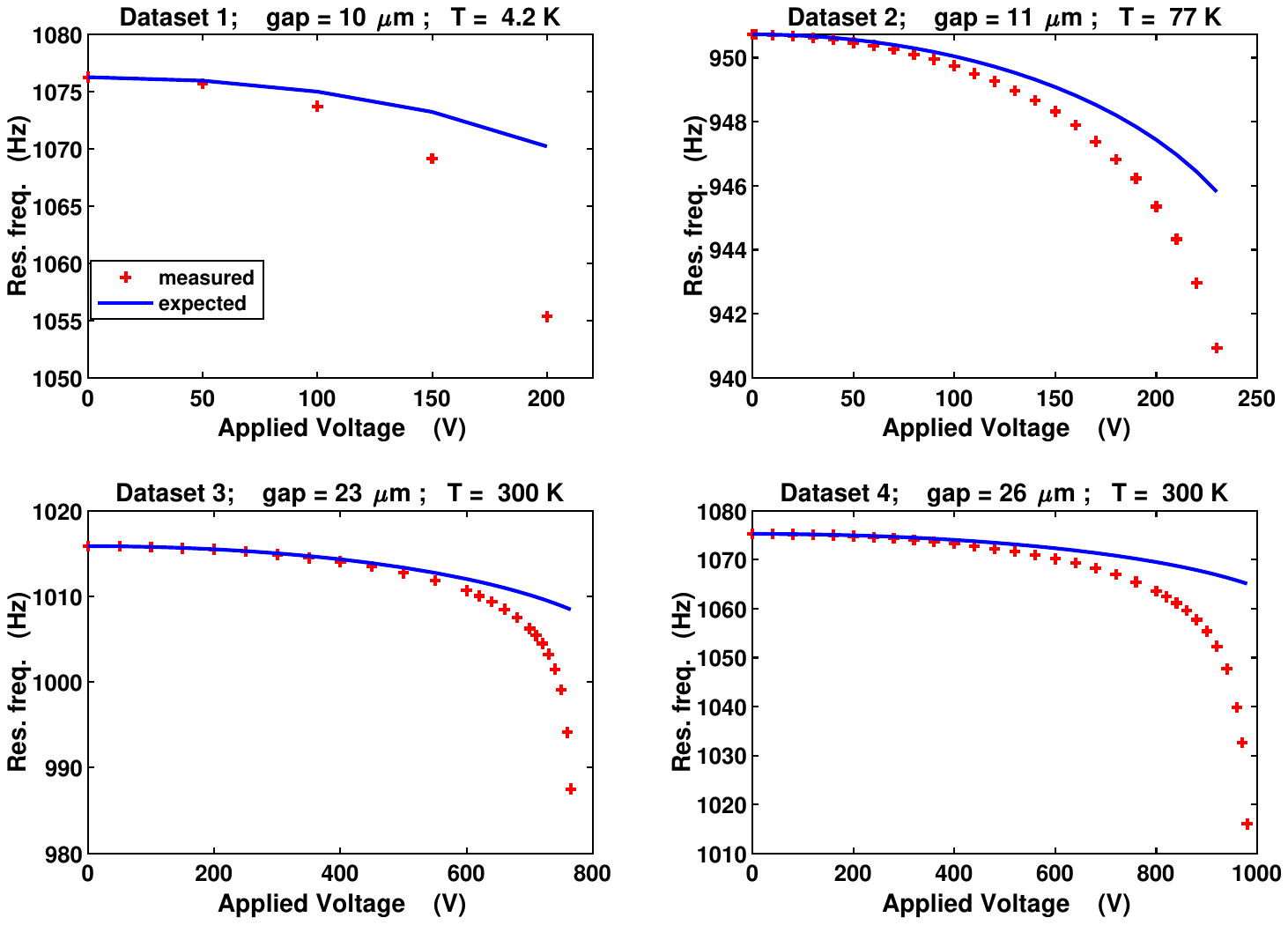}
\caption{Tuning curve for the four experimental set considered. 
Red crosses: measured values; Blue line:  behaviour predicted by eq.\ref{eq:tune}.  Departure from quadratic behaviour is particularly evident in datasets 3 and 4. } 
\label{fig:tuning}
\end{center}
\end{figure}

 However,  in all experimental instances, the frequency change of the resonator, due to the electric field, is much larger than expected, and we must assume an additional attractive force $F_{un}$ giving origin to an additional stiffness $ K_{un}$:
 \be
    f^2_{meas}  = \frac 1{4\pi^2} ~ \frac{K_m - K_{el} -  K_{un} }m
    \label{eq:f2}
  \ee
The result is shown in fig.(\ref{fig:tuning}) for  four of the data set considered, where we show the expected resonant frequency $f_c$, predicted by eq. \ref{eq:tune} and the measured frequency $f_{meas}(V)$.  \\
{\it In all cases, the measured tuning is larger than the  predicted one. Moreover, the experimental values lie on a curve with a sharp knee, a faster-than-exponential behaviour, rather than obeying the expected quadratic law. }

 
\section{The Apparatus}
\begin{figure}[htbp]
\begin{center}
\includegraphics[width= \larghez]{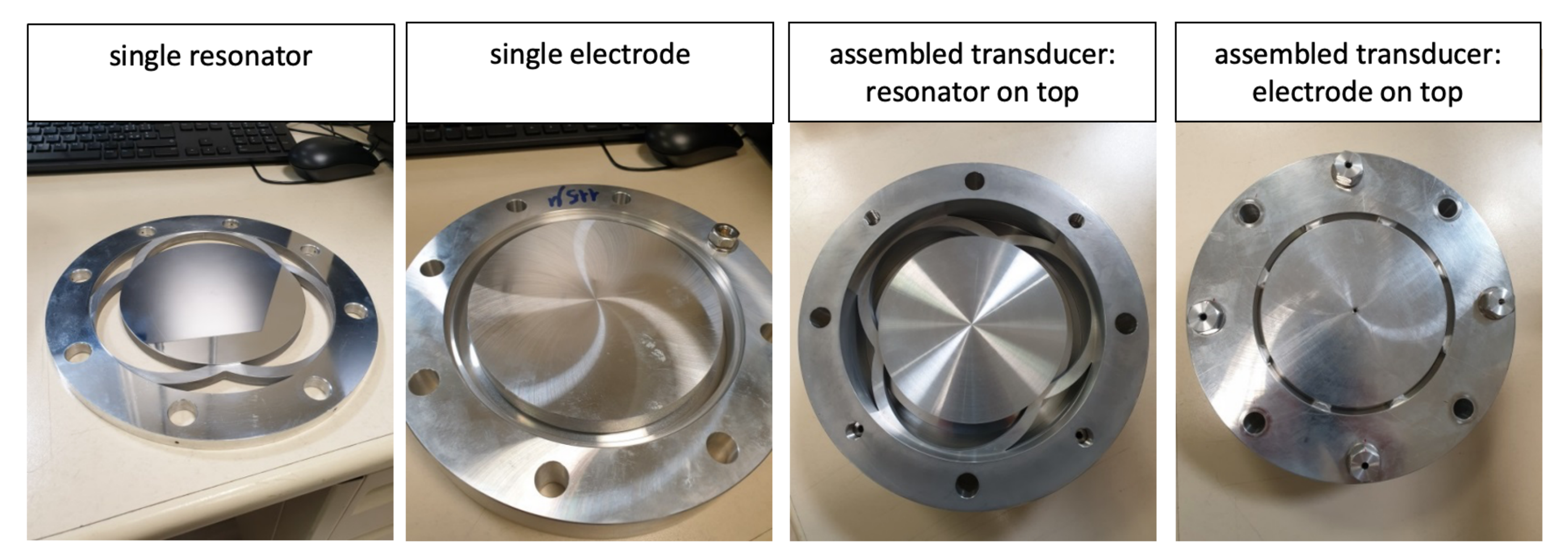}
\caption{Images of the resonator plate, of the  electrode plate and of the assembled transducer}
\label{fig:foto}
\end{center}
\end{figure}

We describe here the devices used  in our six data sets:  they are all 
composed of two\footnote{Three disks  for the push-pull configuration, dataset 6, that has an electrode on each side of the resonator.} disks, shown in fig. \ref{fig:foto}. 
The {\it resonator} is a massive (0.3 to 0.65 kg) disk supported by six curved cantilevers  at the center of a thicker ring. Resonator, cantilevers and supporting ring are e.d.m. machined out of a single slab of Al 5056.
 The supporting ring of the resonator is bolted to  a facing  thick ring containing, at its center,  {\it the electrode},
 isolated by means of PTFE spacers. The spacers' thickness is initially 100 $\mu m$ but, after insertion of the electrode by thermal differential contraction, is squeezed to about 70 $\mu m$.  The assembled transducer is then bolted on a thick (5 cm) solid disk, not shown in the pictures, to ensure that the resonator supporting ring would not take part in the oscillations. The entire assembly is made of the same nonmagnetic alloy Al5056. 

\begin{figure}[htbp]
\vskip-5mm
\begin{center}
\includegraphics[width=.65\larghez]{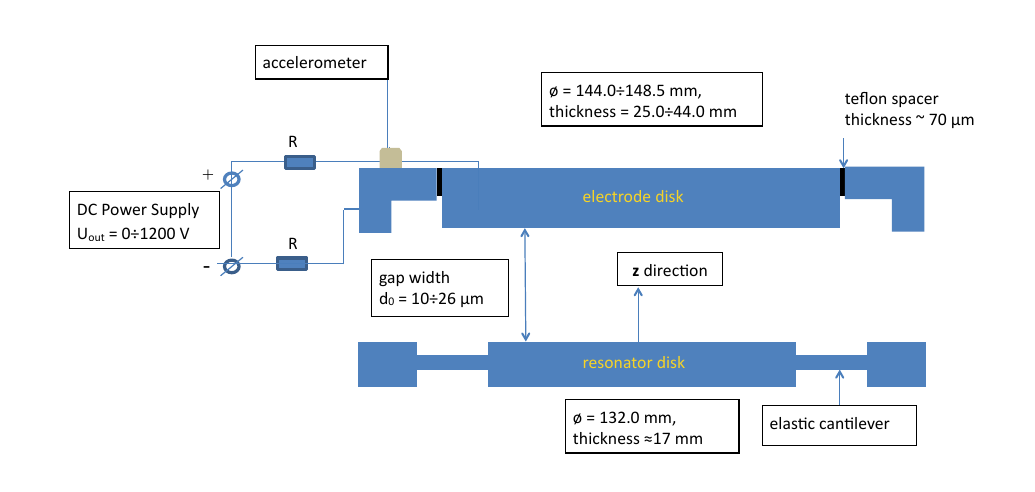}
\includegraphics[width=0.32\larghez]{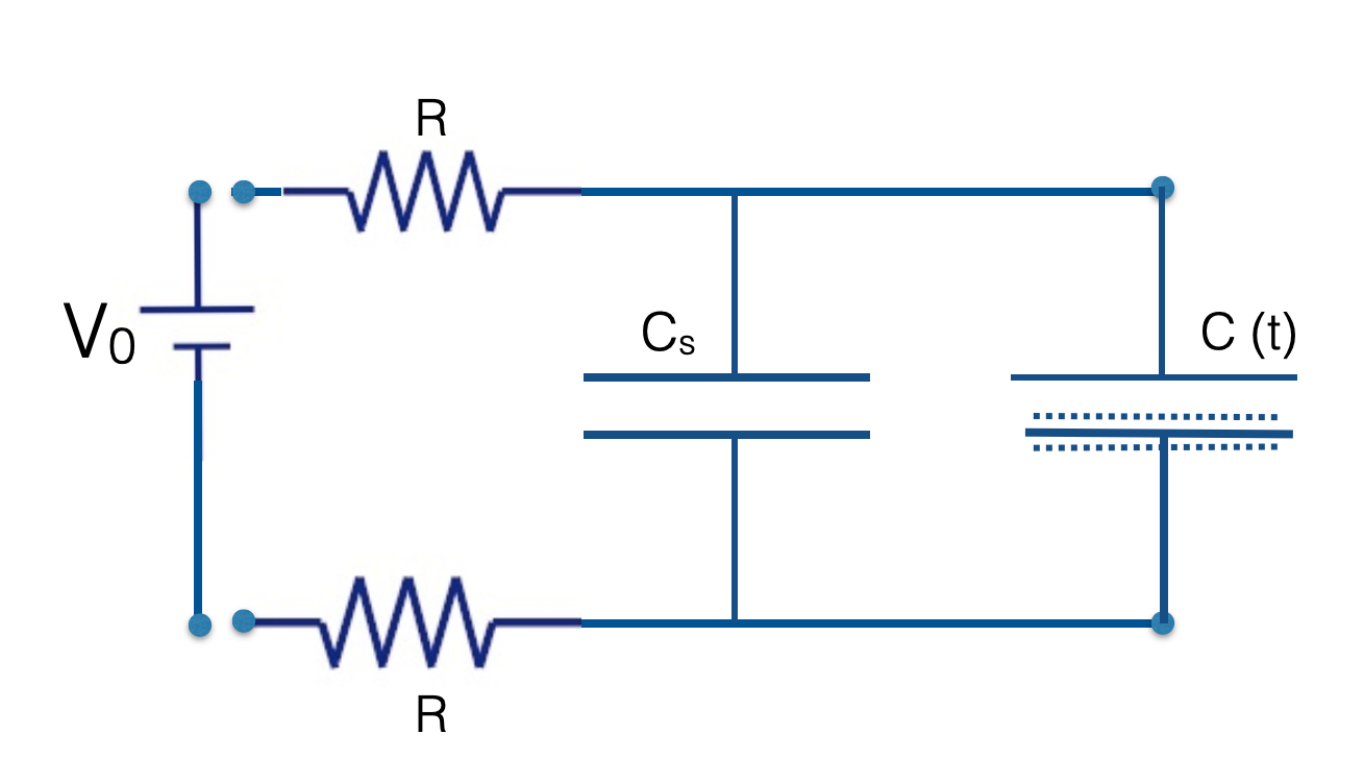}
\caption{Schematics (side view)  and equivalent circuit of the apparatus. The bias resistors varied in the range $ 10^3 - 10^7 $ Ohm. The d.c. voltage source was then disconnected during measurements.}
\label{fig:schema}
\end{center}
\end{figure}

The apparatus is schematically shown in fig.\ref{fig:schema}:  the stray capacitance of the electrode is accurately measured just by moving the resonator away from the electrode plate.
The capacitor is charged to a large potential difference, up to 1000 V, by a voltage generator, through two large (R= $ 10^3 \div10^7~ \Omega$) bias resistors. 
The vibrations are excited, to amplitudes ranging between few nm and fraction of a  $\mu$m, by a piezoceramic  and detected by an accelerometer, bolted on the electrode ring. In this way, excitation and readout do not interfere with the electromechanical set-up of the charged resonator.  The resonant frequency is found either by sweeping the range around the fundamental mode with a sine wave excitation at constant amplitude, or by exciting the PZT with white noise. 
The vibrations are then processed by a digital spectrum analyzer that provides amplitude, frequency and decay time.
Measurements are carried out in vacuum, with a residual pressure varying in the $10^{-2} - 10^{-4}$ Pa range.

We consider  six measuring sets, using four different resonators, at both ambient and cryogenic temperatures. 

\begin{table*}[ht]
\caption{Parameters used in the calculation. Measuring errors are  1\% on capacitances and gaps, on the last displayed digit for all others.  Data of set 4 are used in the following examples.}
\begin{center}
\begin{tabular}{|c|c|c|c|c|c|c|c|}
\hline
&Data set& & set 1 &  set 2 &  set 3 &  set 4 & set 5 / 6\\
\hline
&Device &  &TR\_Naut &TR\_Expl &TR\_Naut & TR\_Expl\_2 & PushPull\\
\hline
\hline
Quantity &Symbol & Units     \\
\hline
T& Temperature &  K& 4.2& 81&  300 & 300 & 300 \\
$R_0$ & resonator radius &mm & 66    &&&&\\
t & resonator thickness  &mm & 17  & 17.35&17 &17.35& 9.6\\
m & resonator mass &kg & 0.625  &0.642 &0.625& 0.644 &0.361\\
$C_0$ & Active capacitance   & $nF$&12.16 & 11.25  &5.24 & 4.68 &6.15 /6.19   \\
$C_{s}$ & Stray capacitance & $ pF$  &700 & 380  &260 & 290 &  450 / 460 \\
$d_0$ & Capacitor gap &$\mu m$&10.0  &10.8 & 23.1 &25.9  & 19.7 / 19.5  \\
$K_{m}$ & Elastic constant & $  MN/m$ &28.6  & 22.9 &28.6 & 29.4   & 11  \\
$f_0$  & Res. freq. @V=0 & $ Hz$ &1076.26 & 950.73  &1015.88 & 1075.29  &  876.42  \\
$V^{max}$  & Max. bias voltage & $ V $ &200 & 230  &765 & 980  & 220  \\
$\alpha^{max}$  & See eq.(\ref{eq:alpha}) @  $V^{max}$ &  &0.084 & 0.114  &0.112  & 0.115  & 0.035  \\
$d^{max} - d_0$  & gap shrink @  $V^{max}$  &  $\mu m$  &-1.0 & -1.6  &-3.3  & -4.1   & \\
\hline
\end{tabular}
\end{center}
\label{tab1}
\end{table*}
We remark that two of these devices had been implemented as resonant vibration transducers in the cryogenic, resonant-bar gravitational wave detectors Explorer \cite{Expl} and Nautilus\cite{Naut}, where they operated consistently and reliably, with no measurable loss of charge, for years.  More details on these ``rosette"  resonant transducers can be found in refs.\cite{trx1,trx3}. \\

\section{Experimental evidence}
\subsection{Simple capacitor, data set 1 to 4}
The data sets 3 and 4, taken at room temperature, extend to larger bias field, approaching the threshold of breakdown.  For data set 4 breakdown  was actually achieved at  990 V.  
{Sets 1 and 2, instead,  were gathered in cryogenic enviroment, where thermal contraction shrinks the gap by several $\mu m$, and prudence had to be  exercised to avoid the risk of electric discharge. \\
Fig.\ref{fig:tuning} clearly shows how, in all mesurements performed, the resonator tuning was much larger than what expected on the basis of the mechanism of sect. \ref{sec:normaltuning}.\\ 
 Any attempt of explaining the extra tuning on the basis of eq.\ref{eq:tune} (e.g. a large, unaccounted for, parasitic capacitance $C_{s}$) can be dismissed by the consideration that the tuning,  at large V,  departs from the quadratic $\Delta f^2$ vs $E$ law predicted by 
 eq.\ref{eq:tune}, but has a much steeper dependence. \\
 Indeed, data sets 3 and 4 clearly show that the tuning behavior is faster than exponential, and the only satisfactory   fit to those curves is achieved with a divergent function of the kind : 
\begin{equation}
\frac{f_c^2 - f^2_{meas}  (V) }{ f_0^2} =   \frac{ A \cdot V^2}{1- B  V }
\label{eq:diverge}
\end{equation}

 where A and B are fitting parameters. In the following, we discuss as an example  the measurements of dataset 4, where the applied voltage extends up to 980 V.\\
 Figure \ref{fig:DeltaK} shows the measured  stiffness change vs applied voltage for dataset n.4: the yellow line shows the amount of extra stiffness $K_{un}$ to be accounted for. 
 \begin{figure}[htbp]
\begin{center}
\includegraphics[width = 0.6\larghez]{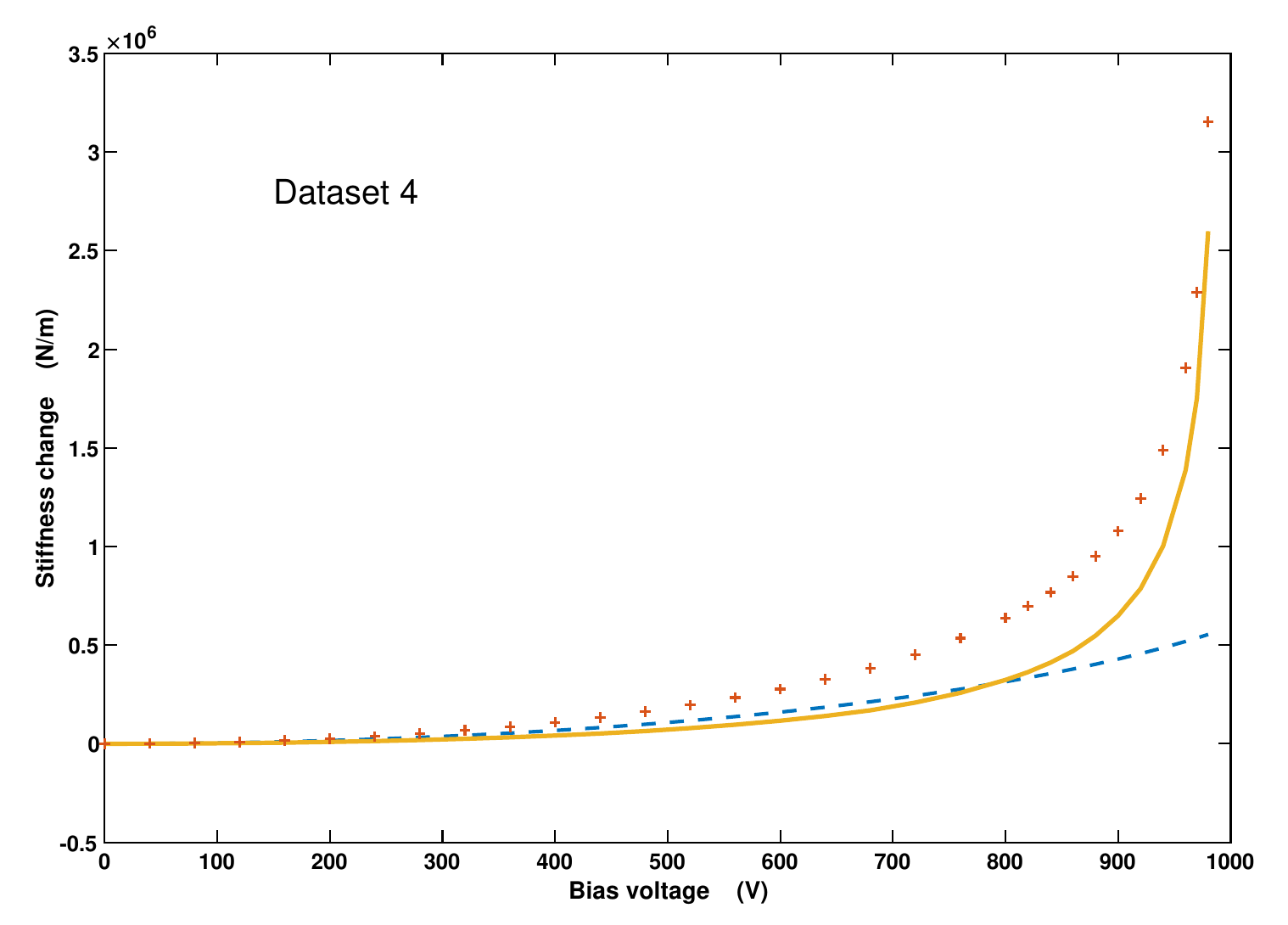}
\caption{Example of anomalous tuning, in dataset 4:   measured stiffness (red datapoints) vs applied voltage. Also shown are the expected stiffness change $K_{el}$ (dashed, blue)
and the unaccounted (solid,yellow) $K_{un}$.}
\label{fig:DeltaK}
\end{center}
\end{figure}

   \subsection{More evidence: ``the push-pull" configuration}
  We have also run experiments \cite{yuri} with the mechanical resonator faced by two electrodes, one per side: this configuration is known as  "push-pull". A push-pull transducer modulates the two gaps with opposite sign. These are usually biased at opposite voltage, thus doubling the signal. We have instead biased the electrodes with identical voltage: this should produce no net electrical force, because two equal extra static forces pull in opposite directions and cancel out.  On the contrary, relevant to our purposes, a double tuning effect takes place: when the plate oscillates, the a.c. components of the electrical stiffness sum up providing twice as much the effect of eq.\ref{eq:tune}. \\
Experimentally, we observe again (fig.\ref{fig:pushpull}) a larger than predicted tuning, and the tuning rate $\Delta f^2 / \Delta V^2$ is, as expected,  double than when only one gap is biased. \\
Unfortunately, the tuning data were only gathered at relatively low voltages, far from near-breakdown: therefore, the anomalous tuning, although well visible, does not show (yet) the diverging behaviour as the other datasets.


   \begin{figure}[h]
\begin{center}
\includegraphics[width=.8\larghez]{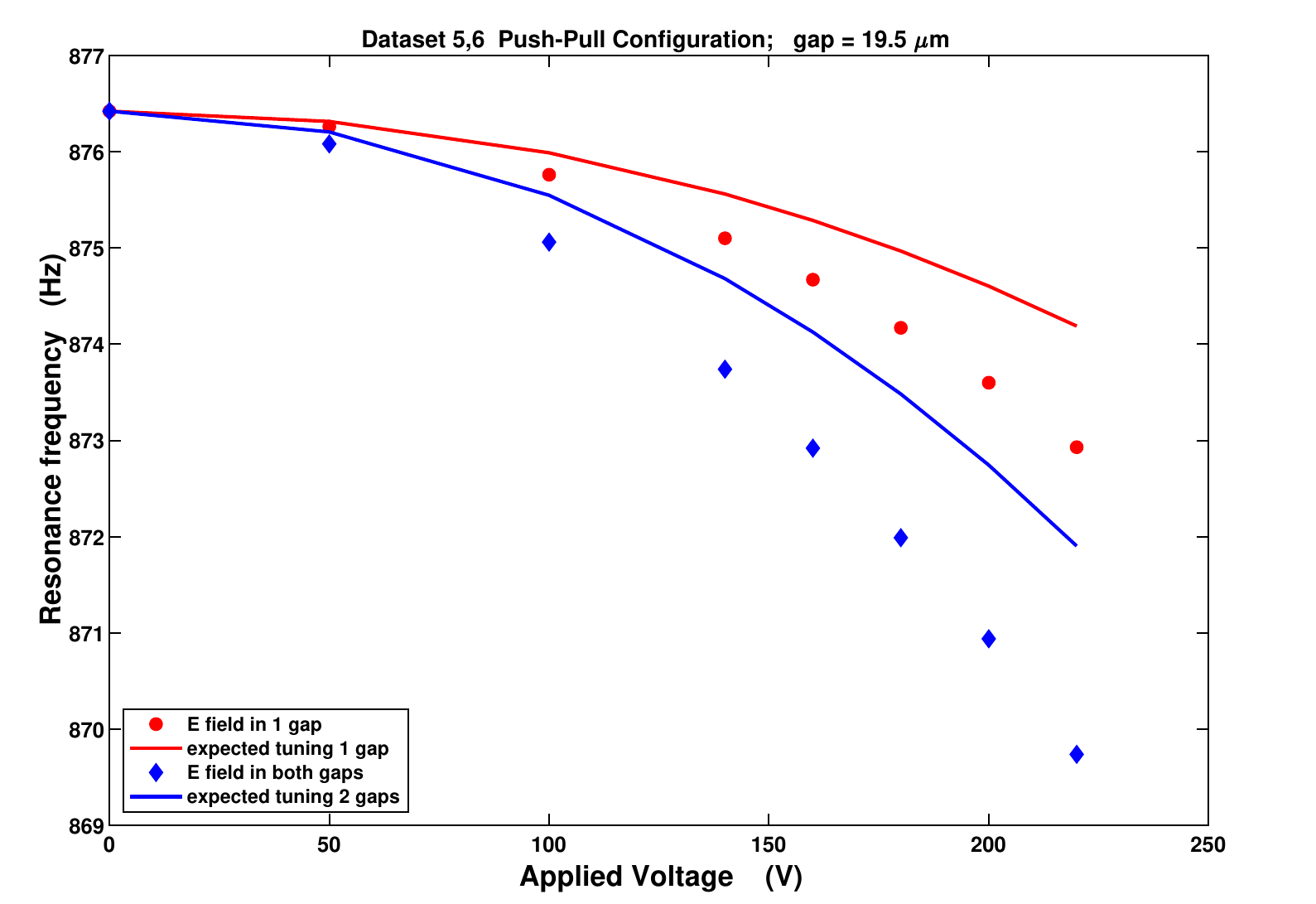}
\caption{Tuning of the push-pull transducer. The plot shows the frequency vs voltage behaviour both  in the case of d.c. bias in one gap (red dots)   and in both gaps (blue diamonds). The expected tuning, based on eq.\ref{eq:tune} are also shown (full lines of respective color) for both experiments.}
\label{fig:pushpull}
\end{center}
\end{figure}

 \section{ Two simple explanations, and why they do not work} 
\label{sec:shrink}

  In the Appendix, several possible explanations for this extra force are considered and discarded.  Here we deal in some detail with the two that most immediately come to mind:  the shrinking of  the gap due to the pull of  the electric force  and the softening of the mechanical restoring force.
\subsection{Shrinking of the gap}
The gap change is a non-negligible effect, and was taken into account in our modeling.  
We start from the geometrical data detailed in fig.\ref{fig:schema}.  The gap $d_0$ is computed from the  measured ``active" capacitance when the capacitor is not charged (V=0).
The gap will change due to the electrostatic attraction of the two plates. The new equilibrium position  $ d_{eq} < d_0$ of the vibrating plate is given by balancing the electric force $\vec F_{el}=  + \frac \sigma 2 S E \hat z$ and the elastic restoring force, $ \vec F_{m} =  -K_{m} (d_{eq} - d_0)\hat z $,  characterized by the spring constant, or stiffness, $K_{m}$.
The new plates separation $d_{eq}$ is thus given by
\begin{equation}
K_{m} (d_{eq}-d_0) = -  \frac{\varepsilon_0  S V^2} {2d_{eq}^2} 
\label{eq:d_eq}
\end{equation}
 that we  rewrite as 
 \begin{equation}
  (d_{eq}/d_0)^3 -(d_{eq}/d_0)^2 + \alpha =0 
\label{eq:d}
\end{equation}
\begin{equation}
\textrm{where } \hskip10mm  \alpha(V)  \equiv \frac{\varepsilon_0  S V^2} {2~K_{m}~d_0^3} 
\label{eq:alpha}
\end{equation}

This is a third degree equation in $d$, that  can be numerically \footnote{Approximate analytic solutions of first and second degree can be given,
 e.g. :
\begin{equation*}
d_{eq}(V) = \frac {d_0}4 [ 3 + \sqrt{1- 8 \alpha}~  ]
\end{equation*}
This quickly shows that no equilibrium position exists for $\alpha > 1/8$. Our datasets 2,3 and 4  (tab.\ref{tab1})  came very close to this maximum allowed bias.}
solved.
The computed value of $d_{eq}- d_0$ is in the range    -1 to  -4 $\mu m$ (see table \ref{tab1}), i.e.  a non-negligible correction to gaps ranging 10 to 26 $\mu m$.  So, we insert the modified gap in our calculation and compute the tuning curve of eq.\ref{eq:tune}  with $E  = V / d(V)$ and $C=\varepsilon_0 S / d(V)$.  In this way we obtain the predicted tuning, blue curves of fig. \ref{fig:tuning}.

One could argue that the pull-in effect of the electric field is larger than what we computed:  \\
indeed, by inverting eq.\ref{eq:tune} we can solve for the gap required to provide the measured tuning: 
for dataset 4, 
at maximum bias, the required gap should shrink from $d_0 =26 \mu$m to $d=9.1 \mu$m.
This is just not possible: it is easily shown that there are no stable solutions of eq.\ref{eq:d} for $ d< \frac 23 d_0$:  for smaller gaps the mechanical force can no longer balance the electrical attraction and  the capacitor collapses (a "mechanical short circuit"). 
Recalling that tuning depends on  stiffness, rather than force (eq.\ref{eq:omegatuning}), we  derive an important consequence from the previous consideration:  to account for the anomalous tuning  we cannot invoke a large attractive force (e.g., by a larger electric field), that would cause collapse of the capacitor, but rather seek a force with a relevant vertical gradient $ \frac{d F}{dz}$.
\subsection{Softening of the mechanical spring}
The second possible objection is  that the mechanical stiffness $K_m$ may not be constant in eq.\ref{eq:d} and ``soften" with increasing E field. In this case,  the equilibrium gap would be smaller than what we computed and the resonant frequency would decrease by a mere mechanical, anelastic effect.
The measurement of $K_m$ is reported in Appendix A and shows a perfectly elastic behaviour of the cantilever beams at least up to $F_{max} = $ 110 N, comparable to the maximum electric force.  This force distributed over the surface of the six cantilevers (roughly 400 $mm^2$ each)  corresponds to a pressure  $ p= F_{max} /S_{cant} = 46  $ kPa,  four orders of magnitude smaller than the yield strength, 345 MPa for AL5056 \cite{yield}.
The possible influence of the electric field on the mechanical properties of the cantilevers is discussed in sect.\ref{sec:stif}.

In conclusion,   an extra, unexpected force is needed to account for the observed tuning of all resonator.

\section {Investigating the extra force}  
\subsection{Modeling the additional stiffness}
Let us summarize a few points of what discussed so far: \\
- We have evidence of the presence of an additional restoring force 
that gives rise to an anomalous negative stiffness $K_{un}$.\\
- In Appendix B we list and discuss several physical mechanisms that we have considered, and ruled out, as possible sources of the additional force $F_{un}$.  \\ 
- This extra  force   depends  on the applied electrical field, and has a diverging dependence,  well described by eq.\ref{eq:diverge}.  \\
-  The ``regular" electrostatic force of attraction is  $\vec F_{el}=   \frac { \varepsilon_0~\varepsilon_r S  V^2}{2 d_{eq}^2} \hat z$.
As discussed in sect. \ref{sec:normaltuning}, the force develops during the charging of the resonator, and must thus be computed in constant voltage conditions.\\
Considering that both the capacitor surface $S$ and the bias voltage $V$ are fixed, and the gap $d_{eq}$ can't be shrunk to less  than $2 d_0/3$, as stated in sect.\ref{sec:shrink}, the only handle we have to increase the force,  if we want to remain within an electromagnetic framework, is the dielectric permittivity $\varepsilon_r$. 
A larger permittivity would increase the charge in the gap, and hence the force between the plates.
We postpone to the next section speculations on the origin of an  $\varepsilon_r > 1$,that anomalously increases with the applied field.
 We recall here that anomalous increases of  permittivity are well known phenomena that can occur in dielectric mixtures \cite{khrapak}, under insulator-conductor phase transitions when the density of dielectric inclusions increases. Description of such phenomena gave rise to the Clausius-Mossotti equation, but also to the  Maxwell-Garnett and other similar mixing formulas  \cite{mixing}.\\
Taking inspiration from these effects, we try and analyze our data with a similar approach. However, as nothing is known about the dielectric substance involved in this process, we proceed with a time-independent, mean field approximation, relying on the universality of such method. 
\begin{figure}[h]
\begin{center}
\includegraphics[width=.6\larghez]{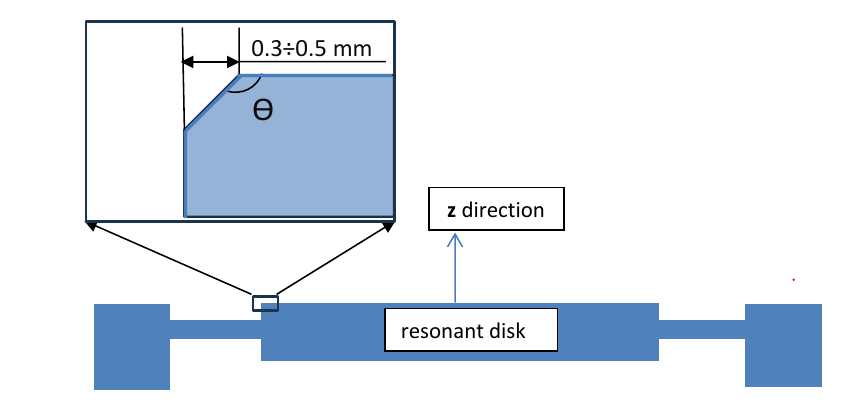}
\caption{A close-up sketch of the edge of the resonator plate, showing the chamfer at angle $\theta = \frac 34 \pi$ }
\label{fig:edge}
\end{center}
\end{figure} 

- We  recall here an additional piece of information  \cite{LL7,yuri}: near the  edge of a parallel plate capacitor,    the electric field has a larger value and non-uniform behaviour:
\be
E_{edge} = E~ n~ \sqrt 2~\sin( \frac{\pi n} 2)  ~\bigg( \frac {d}{R_0} \bigg)^{n-1}  
\label{eq:Eedge}
\ee
where $E = \frac {V }{d_{eq}} $ is the electric field in the rest of the capacitor;  $d$ is, from now on, a shorthand notation for $d_{eq}(V)$; 
$ n \equiv (2 - \theta/\pi)^{-1}$ and $\theta$ is the internal angle of the edge of the resonator plate {(see fig.\ref{fig:edge}).
 In our case, 
$\theta = \frac 34 \pi$  so that n = 0.8.  Moreover, we can drop the factors  $n~\sqrt 2~ \sin( \pi n/ 2)  \simeq 1$.
This enhancement of the electric field, of a factor $\sim$ 5  
extends  over an annular surface $\Delta S~$ of width comparable, as order of magnitude, with the gap size. 
\begin{figure}[h!]
\begin{center}
\includegraphics[width=.8\larghez]{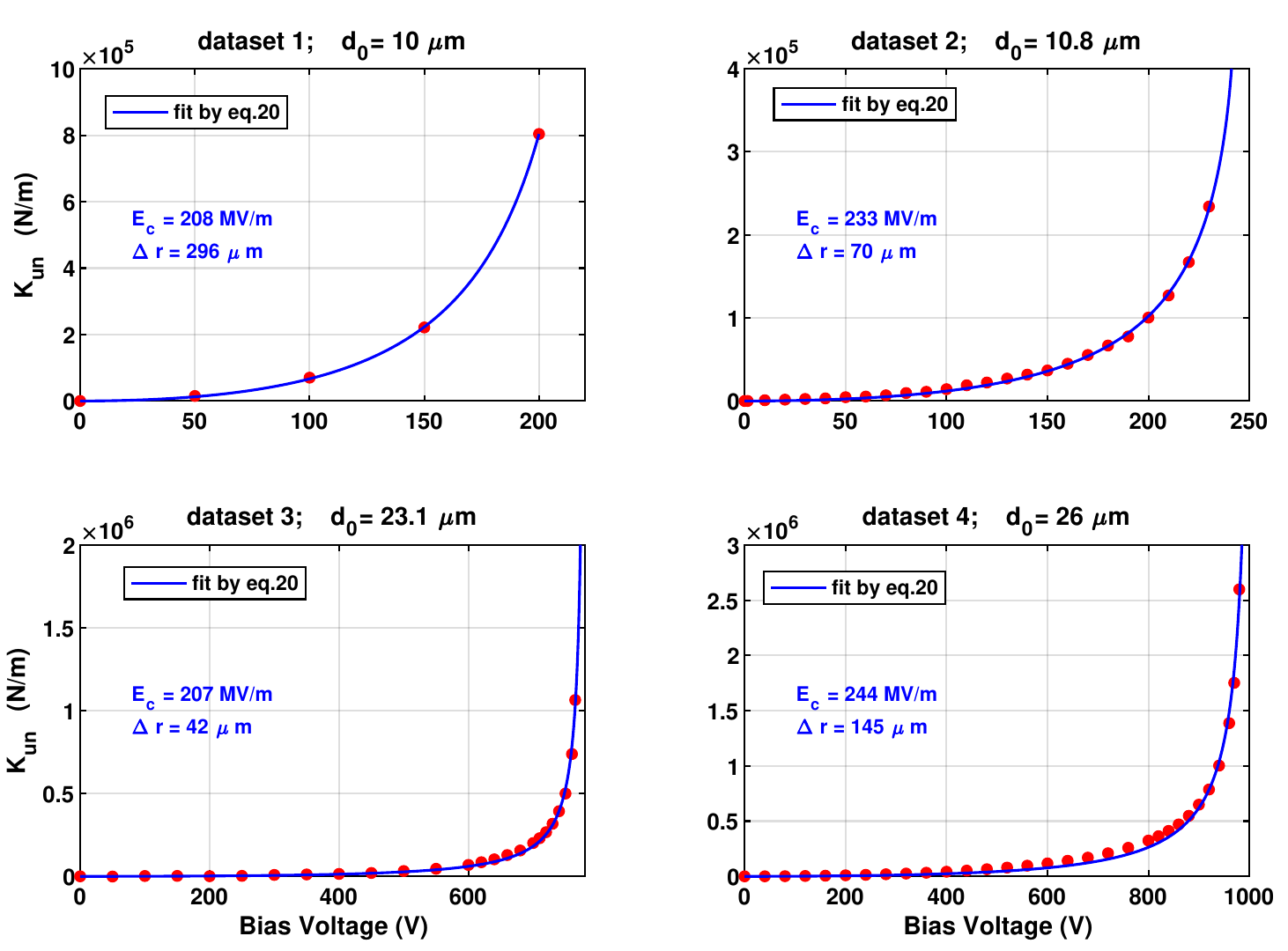}
\caption{The extra stiffness $K_{un}$ as derived by tuning measurements in datasets 1-4. The data are fitted by eq. \ref{eq:Kunk2}, yielding the fitting parameters shown in the plots and summarized in table \ref{tab:fit}. Measurements for datasets 5 and 6 are too far from critical region and do not yield unambiguous results }
\label{fig:fit}
\end{center}
\end{figure}

 We  assume that  the field facing this region $\Delta S$,  a fraction of a percent of the whole active surface, becomes large enough to activate the runaway process shown in fig.\ref{fig:DeltaK} 
and allows us to write:  
\be
\chi_{edge} = \varepsilon_r -1 \equiv  \frac 1{1 - E_{edge} /E_c} 
\label{eq:chi}
\ee
where $E_c $ is a critical field\footnote{This form provides a good approximation close to the critical point. Although it  does not offer the correct limit $\chi \rightarrow 0$ for $E \rightarrow 0$, the factor $E_{edge}^2$ in eq.\ref{eq:energy2} below guarantees the correct behaviour }.
The potential energy of the charged capacitor is thus:
\be 
W_{el} = W_{disk} + W_{edge} \simeq  \varepsilon_0 \frac{S d} 2 E^2 + \varepsilon_0  \chi_{edge} \frac{\Delta S~ d} 2 E_{edge}^2
\label{eq:energy2}
\ee
The  total electric force acting on the vibrating plate is found, once again, by taking
the derivative of $- W_C$ with respect to the plates distance $d$.  By plugging in the above expression for $E_{edge}$   we modify eq.\ref{eq:d_eq} to include the extra force

\be
F_{z} =  K_{m} (d_{eq}-d_0) +\varepsilon_0 \frac{S } 2 E^2 + \varepsilon_0  \chi_{edge} \frac{\Delta S~ } 2 E_{edge}^2~ (2n -1)
\label{eq:Fc}
\ee

  This is the electrostatic force we should use, in place of eq.\ref{eq:d}, to find the equilibrium gap.  \\
 Moreover,  eq.\ref{eq:Fc} yields an additional stiffness:
\be
K_{un} = \frac {\partial F_{un}}{\partial z} =  \varepsilon_0  \chi_{edge} \frac{\Delta S~ }{ 2 d} E_{edge}^2~ (2n -1)(2n -2)
\label{eq:Kunk} 
\ee
Finally, inserting  eq.\ref{eq:Eedge} for $E_{edge}$,  $2 \pi R_0 \Delta r = \Delta S$,  and  eq.\ref{eq:chi}, with its divergent behaviour,  for $\chi_{edge} $, we obtain:
\be
K_{un} =g_1  \frac { E_{edge}^2}{1 - E_{edge} /E_c}~ \frac{ \Delta r}{d} 
\label{eq:Kunk2}
\ee
where 
\footnote{ see footnote at eq.\ref{eq:Kel} about the sign of $K_{un}$.} $ g_1 = R_0 \pi  \epsilon_0  (2n-1)| (2n-2)|  \simeq  4.41~10^{-13}~F $ 
 and $E_{edge} $  as in eq.\ref{eq:Eedge}.
 
Fig.\ref{fig:fit} shows how well   eq.\ref{eq:Kunk2}  fits the measured extra stiffnesses, as defined in eq.\ref{eq:f2}: 
$$K_{un}  = [1 - (f/f_0)^2]\cdot K_m  - K_{el}    
$$
 in all datasets.  
The  least-square fit provides the two fitting  parameters   $E_c$ (critical field)  and  $\Delta r$ (edge  radial extension), that are shown in 
table \ref{tab:fit}.

\begin{table}[htp]
\caption{The critical field $E_c$ and the radial extension $\Delta r$ of the edge region where the electric field enhancement (eqs.\ref {eq:Eedge} and \ref{eq:chi})  should take place. These values are derived by fitting the experimental values for $K_{un}$ with eq.\ref{eq:Kunk2}.  Row 5 reports the shrinked gap $d_{eq1}  $, at maximum bias voltage,  due to the "conventional" electrostatic, force eq.\ref{eq:d_eq},  while, in row 6, we compute $ d_{eq2}$ also accounting for the additional term of eq.\ref{eq:Fc}.  }
\begin{center}
\begin{tabular}{|c|c|c|c|c|c|}
\hline
               Dataset &   1     & 2     & 3     & 4     \\ \hline \hline
$E_c~~$         (MV/m)           & 208 & 233 & 207 & 244 \\ \hline
$\Delta r~ ~$  $(\mu m)$        & 296 & 70  & 43 & 147 \\ \hline  \hline
$d_0 ~~~$        $(\mu m)$        & 10.0  & 10.8  & 23.1  & 25.9  \\ \hline
$d_{eq1} ~~$    $(\mu m)$        & 9.0   & 9.1   & 19.6  & 21.8  \\ \hline
$ d_{eq2}~~$    $(\mu m)$      &  8.5   & 8.9   & 18.9  & 20.5 \\ \hline 

\end{tabular}
\end{center}
\label{tab:fit}
\end{table}

The values of $\Delta r$  provided by the fit, and reported in table \ref{tab:fit}, show that the edge effect does extend over a region of the order of few gap widths.  
It appears as if  $E_c \sim( 2.2 \pm 0.2)  \cdot 10^8$ V/m is a critical field of the same physical process for all datasets, regardless of gap width or temperature.    When the  edge field reaches, e.g., $0.9~E_c$,  the factor $\chi_{edge} $ increases to $\sim $ 10, so that the energy density  in the edge can grow (eq.\ref{eq:energy2}) to $w = 2~ MJ/m^3$.

\subsection{A daring suggestion about the origin of the extra force}
We mention here a far-fetched idea that could lead to an increased $\varepsilon_r$ through vacuum fluctuations.\\
As discussed and shown in the previous section, the charged capacitor  behaves as if its gap were filled with a dielectric mixture  
where the  the concentration of dielectric inclusions increases with the electric field. Only three dielectrics are present in the devices: PTFE spacers insulating the electrode, a small amount of residual gas (mostly air at reduced pressure, $p \sim 10^{-3}$ Pa)  and physical vacuum in
the transducer gap.\\
We can exclude a role of both the PTFE spacers (sect. \ref{sec:ptfe}) and the residual gas (sect. \ref{sec:gas}): for both, the permittivity remains constant up to the largest applied fields.

So, we are left with only physical vacuum. Can vacuum exhibit properties that can cause  macroscopic effects
for the electric charge in the gap ?  
Physical vacuum can be considered as a dielectric many-body quantum system 
 because of the possible presence of various types of vacuum fluctuations,
including virtual electron-positron pairs, as  quantum oscillators in their ground state, of different frequencies: this implies different energies  $\hbar \omega$, dimensions, lifetimes.

The classical tests of QED \cite{Lamb,Kusch,Gabrielse}   proved that
virtual electron-positron pairs surrounding a charged particle,  at  distances comparable with  the Compton length, can be polarized by the strong electric field of this particle, thus creating  induced virtual electric dipoles. An extreme case of vacuum polarization is the Schwinger effect \cite{Schwinger}, i.e.  the creation of real electron-positron pairs in extremely  strong  ($\sim 10^{18}$ V/m) electric fields. The issues of vacuum fluctuation polarizability and interactions were studied in \cite{Fedorov}.
 Virtual pairs of lower energy,whose volume concentration is much less, have much larger polarizabilities and longer lifetimes (according to uncertainty principle $ \Delta \omega \Delta t \sim 1$). This allows these virtual pairs to be significantly polarized   by our relatively low  electrical field, throughout the entire volume of the capacitor gap  and providing sufficient time to interact with the charged surfaces so to increase, during the charging phase, the net elctrode charge. This cannot take place with the very weakly polarized, very short-lived higher energy density fluctuations. 
 Significant polarization occurs when the energy density of a virtual pair is comparable or less than those of the gap electrical field: at $E_{edge} \sim  E_c$ it corresponds to energy densities $\le 10^6  J/m^3$ and lifetimes $\ge 10^{-14}$ s. 

In this way, the vacuum in the gap can be considered as a dielectric mixture of an homogeneous environment with $\varepsilon_r^{vac}= 1$ and dielectric inclusions presented by polarizable long-life virtual pairs. The resulting effective permittivity of the mixture can be described by the conventional mixing formulas. Since the initial (at low voltage) concentration of such long-living virtual dipoles should be very small, the question is: can their concentration increase significantly with the  electrical field in the gap ? 

We try here to suggest simple considerations that can account for our data.
Some properties of polarized  and non-polarized virtual pairs could be quite different. A virtual dipole may have larger volume, lower frequency and therefore higher lifetime. The increase of the lifetime of virtual particles was studied in \cite{Lyubo}. Besides, long-life virtual dipoles can be correlated and interact with each other (dipole-dipole interaction) forming virtual quasi-bound  correlated states between immediate neighbors. This can be considered as the formation of a new single object of higher virtual dipole concentration, higher volume, higher dielectric susceptibility and higher lifetime: we can call it a virtual cluster. \\
Recent experiments \cite{corr1, corr2}
have shown that external electromagnetic fields can induce non-local correlations among vacuum fluctuations. The same effect can be obtained by suitable boundary conditions, like mirrors in a superconductor cavity \cite{cavities} or gap electrodes in a plate-parallel capacitor . \\
Both correlation and extended lifetime should contribute to  increase, with increasing electric field,  the concentration of  virtual dipoles, forming virtual clusters and triggering a possible self-sustaining process: a larger volume increases the lifetime, the longer lifetime increases the probability of new bound states which increases the volume and so on, like a condensation of virtual dipoles around the core of a new phase. \\
Such hypothetical process resembles an insulator-conductor phase transition under the influence of electrical field where virtual clusters are like critical fluctuations of a new quantum phase of larger dielectric susceptibility. This phase transition ends at some critical point  $E_c$  which represents an upper limit for electrical breakdown. 

These are very simple and speculative considerations, using analogies with physical processes in
condensed matter. In any case, physical vacuum remains a strong candidate for the role of a dielectric
in the gap.

\section{Conclusions}
We have analyzed  tuning data on simple physical systems:  electromechanical devices comprising large, parallel plate, capacitors with a small gap (10 to 26 $\mu m$), composed of a fixed plate and a vibrating  one, biased with a d.c. voltage $V$.  The stiffness of the resonator (measured via the resonant frequency at constant charge, $ K_{meas}= m~\omega_{meas}^2$) is known to have a negative contribution $ K_{el}$ from the electric field in the gap, proportional to $E^2$. However, the measured stiffness $K_{meas}$  is much smaller than the predicted one in all devices, at all voltages, all temperatures. Moreover, the functional dependence on the applied voltage (or on the stored $E$ field) exhibits a divergent behaviour with the electric field.  Therefore we are lead to hypotize that an  additional force $F_{un}$ exists, giving rise to an anomalous additional stiffness $K_{un}$. Several possible explanations of this effect were considered and dismissed (sect.\ref{sec:shrink} and Appendix B). \\
The observed anomalous force is compatible with the presence of a dielectric mixture in the gap, undergoing an insulator-conductor phase transition where the concentration of dielectric inclusions increases with electric field, a process that can be described by Clausius-Mossotti or other similar mixing equations. Recalling that the electric field is enhanced at the edge of the plate, and applying mean field approximations, we find that this approach well describes our experimental data:  all stiffness measurements are fitted by a simple function, eq.\ref{eq:Kunk2}, with a value of critical electric field $E_c \simeq 3 \cdot 10^8$ V/m, nearly constant across all analyzed datasets. 
%
In the gap of each capacitor there is only physical vacuum (residual gas was ruled out, see App. B) which can be viewed as a dielectric mixture where dielectric inclusions are polarizable virtual pairs of low energy density. We dare to suggest that polarized virtual pairs act as induced interacting dipoles, in such a way as to increase their concentration with increasing electrical field and therefore to increase very sharply (as under an insulator-conductor phase transition) the effective dielectric permittivity of vacuum in the gap. This would happen mostly in the edge region, where the electrical field is non-uniform and five-fold larger. 
We hope that these experimental results will foster the development of more advanced models of physical vacuum and stimulate new ideas for more sophisticated experiments in this field.
%

Summarizing, we have the following unexpected, interesting results:  \\
- an additional force acting on sub-millimeter distances has been detected in charged parallel-plate capacitors in vacuum; \\
- there is  evidence that a dielectric mixture-like substance is present, mainly in the edge regions of the gap,  and undergoes an insulator-conductor phase transition under the influence of the electric field, giving rise to this anomalous force; \\
- such unknown dielectric may be physical vacuum itself, considered as a dielectric mixture where the dielectric inclusions consist of efficiently polarizable vacuum fluctuations.


%
\vskip10mm

\section*{ Appendix A - Measuring the mechanical stiffness $K_{m}$}
\begin{figure}[h!]
\begin{center}
\includegraphics[width=0.6\larghez]{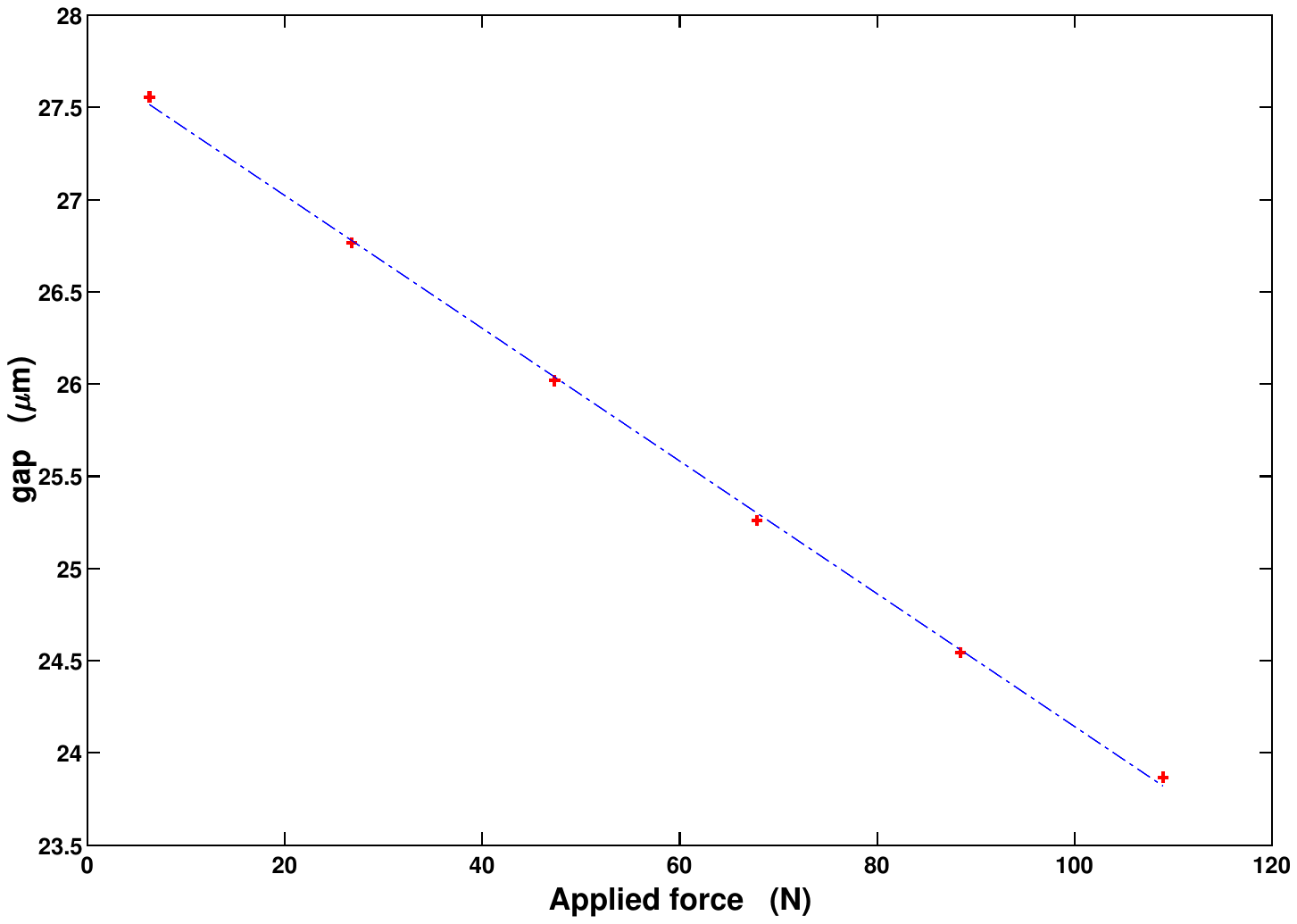}
\caption{Linearity of the gap vs  force relation. The slope of the linear regression, $-K_m^{-1}$, provides the mechanical stiffness.}
\label{fig:gap}
\end{center}
\end{figure}
The spring constant $K_{m}$ characterizes the resonator and is  used in sect.\ref{sec:intro}. We  evaluate it in two ways:
\vskip-10mm
\begin{itemize}
\item  a)  dynamically, from the fundamental mechanical resonant frequency of the oscillator:  
$(2 \pi f_0)^2 = K_{dyn} /m$
where $f_0$ is the starting (at V=0) frequency of the resonator. 
\item b) statically, by loading the resonator with a series of well known load forces $ F_i = g M_i$ (2 to 10 kg) and measuring the change in capacitance due to yielding of the cantilevers.
We find a linear dependence:
$\frac{ \Delta C}{\Delta F_i} = 6.6 pF/N $.
Knowing the capacitor surface, this is easily translated into a weight-displacement relation: 
$$d_i = \frac{\varepsilon_0 S}{C_i} = d_0 - \frac {F_i} K_{stat}  $$
The result, fig.\ref{fig:gap}, shows that the linear relation $ F = - K_{stat} ~d$ holds at least up to 110 N of load, comparable with the maximum force exerted by the electrostatic field. No deviation from Hooke's law is observed and the measurements are very repeatable, excluding any non-linear behaviour or plastic deformation of the cantilevers.
\end{itemize}

With the data of set 4,   the two methods yeld: \\ 
 $K_{dyn} =29.4 $ MN/m ~~ and   $\hskip 5mm   K_{stat} =  28.0   $ MN/m.  The two evaluations agree within  $\sim 5\%$.\\
 A similar, but smaller ($0.5 pF/N $ ) effect takes places when weigth-loading the facing electrode, probably due to sliding of the central disc on the PTFE spacers.

\section{Appendix B - Concurring effects}
We list here a number of physical processes that we considered (and excluded) as possibly contributing to the anomalous tuning.
\subsection{Pressure and temperature dependence}
The tuning data plotted in fig.\ref{fig:tuning} are representative of many data-takings, collected in various enviromental conditions.\\
{\bf Pressure:} the pressure of the vacuum chamber was always below $10^{-3}$ Pa. At all pressures below 1 Pa, the  pressure has no measurable influence on the resonance frequency. 

{\bf Temperature:} Set 1 and 2 were collected at 4.2 K  and 77 K respectively, with temperature  stabilized by the boiling cryogenic liquid.  Set 3 and 4 at room temperature,  $ T =  297 \pm 3$ K, and were repeated many times, with all measurements consistent to within 10 mHz, i.e.  to 1 part in $10^5$: error bars are not appreciable in the plots.
 The one exception is the value at highest voltage in Set 3: with the electric field near breakdown,   the frequency was not fully repeatable and the data point is an average over several measurements.  
 
 \subsection{Oxide and moisture}
According to literature \cite{oxide}, a polished Al surface develops, when exposed to air, an oxide ($Al_2O_3$) layer of thickness 1-10 nm. This is at most 1/1000 of the gap width and, with a permittivity $\varepsilon_r = 9$ can account for less than 0.1 \% of the total capacitance. Moreover, it is a fixed correction, that cannot explain the divergent behaviour of the extra stiffness $K_{un}$.\\
Presence of moisture in the gap  cannot be excluded at room temperature, but it would certainly not be found at cryogenic ( 4.2 K and 81 K)  temperatures.  We underline that a similar anomalous tuning was observed when the transducer of datasets 1 and 3 was operated, on the Nautilus g.w. detector, for many years under continuous cryopumping at T=1.5 K.
 
 \subsection{Irregularities of facing surfaces in the gap}
 The surfaces of both the resonator and the electrode are machined flat and polished to sub-micrometer accuracy. The surface quality was verified on a measuring station.  Assume nevertheless that one spot, of surface  $ S_{spot}\sim mm^2$, sticks out to a smaller distance $d_{spot}$ from the facing surface: it forms a capacitance $C_{spot}$ that adds in parallel to the rest of the surface:
 $$ C_0 = \varepsilon_0 \bigg[\frac{(S-S_{spot})} d + \frac{(S_{spot})} {d_{spot}}  \bigg]$$
 The value of $C_0$, or $C(V)$, that includes the effect of such imperfection,  is directly measured, and it is the quantity that determines the tuning, as shown in eq. \ref{eq:tune}.  Should such a spot exist, it would represent a second capacitance, in parallel with $C_0$: we can reasonably expect that each of the two capacitors would contribute its own $\Delta K$, and these would add.   In order to account for the extra stiffness,  about 700 \% of the ``regular" one at maximum field, we should have 
 $ \frac{(S_{spot})}{d_{spot}} \ge  7\cdot \frac{(S-S_{spot})}{d} $. We can surely exclude that such a spot exists.

\subsection{ Imperfect parallelism of the electrode surfaces}
In this case,  the assembly of the two surfaces  lacks parallelism, so that the gap at one end is $\Delta d$ larger than at the other.  We would have an (infinite) number of capacitors in parallel, for each distance $d(x)$. 
 We run a simpler, worse case calculation for a rectangular plate, and obtain:
$$ C_{wedge} = \varepsilon_0 S \frac 1{\Delta d}~ log\bigg( \frac{d_0 +\Delta d/2}{d_0 -\Delta d/2}\bigg) $$

Let us assume a difference in the gap height $\Delta d = 10 \mu m$  across the resonator width, 
i.e. equal to 40 \% of the gap itself: this is way larger than our metrology allows. Nevertheless, the resulting capacitance $C_{wedge}$ would differ from $C_0$ (perfect parallelism) by:  $ C_{wedge}/ C_0 =  1.013$. Such a tiny difference in capacitance  cannot be the cause of the very large (700 \%)  effect on $ K_{un}$. 
Moreover, the  effect is much smaller when computed for the actual disk geometry, where the line of minimum distance collapses to a point.

\subsection{Capacitance fringe effects}
The contribution to the total capacitance from fringe effect is evaluated  as described in \cite{bordo}. Defining $x \equiv d_0/t$, where $d_0$ is the gap and $t$ the plate thickness (see table 1):
\begin{align}
C_{fringe} = &111.3\frac{R_0}{4 \pi} [log(16 \pi  R_0 /d_0) -3 + (1+x) log((1+x) \nonumber \\
  & - x log(x)] ~~~pF
\end{align}
 it accounts, at most, for  5pF, i.e. one part per thousand of the active capacitance: also this effect can be safely ignored.
 
 \subsection{Casimir effects}
 \label{sec:casimir}
The Casimir effect between two parallel (uncharged !) plates is given by the well known relation  
$$ F_{Cas} = S \frac{\hbar c \pi^2}{240 d^4}  = 18 \mu N \bigg(\frac {1 \mu m}d \bigg)^4$$ 
 Our maximum value for $F_{Cas}=1.8$ nN is therefore  completely negligible  with respect to the electrostatic force attracting the two plates: $F_{el} = \frac12 \sigma S E \sim 100$ N. \\
However, we are interested in the stiffness that this force can provide:
$$ K_{Cas} = \frac{\partial F_{Cas}}{\partial z}=-4 S\frac {\hbar c \pi^2}{240} \frac 1{d^5}   $$
In our case (for $d=10 \mu$m) we obtain $ K_{Cas} = -712 \mu$N/m, compared with $K_{mech} = 28 $MN/m, or $K_{un} \sim 0.3- 1$ MN/m:  a mismatch of 10 orders of magnitude.

 \subsection{Patch effect}
Patch effect is a phenomenon of spatial and temporal fluctuations of the electrostatic surface potential of real conductors, even when the bulk is equipotential. These patch potentials can cause the patch electrostatic force between two parallel conducting plates that can reach a magnitude comparable to the Casimir force at several microns separation \cite{patch1, patch2}. As discussed in sec.\ref{sec:casimir},  such forces are many orders of magnitude weaker than those observed in our experiments. 
These voltages are of the order of  0.1 V maximum, and develop on small areas. This is clearly negligible with respect to our bias voltages, also considering that the electric force scales quadratically with the voltage. 

\subsection{Electromagnetic issues}
\label{sec:emIssues}
The apparatus operates in constant charge conditions, and  the electric field in the gap  $E= \sigma/ \varepsilon_0$  changes as a consequence of charge flowing from the active capacitance C to the stray capacitance $C_s$. 
We evaluate the relevant e.m. quantities with a simple, first order calculation: 
 we assume a harmonic vibration of the resonator modulating the capacitor gap $d(t) = d_{eq} + z ~cos (2\pi f t) $,  so that the capacitance varies as $C(t) \simeq  C_0 \bigg(1 - \frac{z(t)} {d_{eq}} \bigg) $ 
 and we indicate  with the suffix "0"  all quantities as measured at rest. Numerical estimates are given  for a (very large) vibration amplitude z = 100 nm and for the larger bias voltage $V_0$ of  our six  datasets.
$$V(t) =  V_0 \bigg(1 +  \frac{C_0}{(C_0 + C_s)} \frac{z(t)} {d_{eq}} \bigg) ; $$ 
$$ E(t) = E_0 \bigg(1 -\frac{C_s}{(C_0 + C_s)}\frac{z(t)} {d_{eq}} \bigg)  $$
 The a.c. voltage across the capacitor varies between 0.7 and 2.2 mV  and the a.c. electric field between 40-120 V/m for our six datasets, to be compared with 11-29 MV/m of the  d.c. field $E_0$. \\
$$I   =  \dot Q(t) =  2 \pi f \frac{C_s}{(1 + C_s/C_0)}  E_0 z(t) ; \hskip12mm $$
The current flowing from the resonator to $C_s$ and returning on the mating electrode ranges $3 - 10 \mu$A. The force $F_I$ arising from this current  was (over) estimated considering $I$ as flowing in six streams via the six cantilevers. The resulting force is 5-24 pN, i.e., totally negligible. \\
 The time-dependent electric field gives rise to an induced magnetic field, laying parallel to the capacitor surfaces.
It is easily evaluated, neglecting edge effects on the $E$ field, as: 
$$B(r,t) = \frac{V_0 ~2 \pi f}{2 d_{eq}^2 c^2}\frac{C_s}{C_0 + C_s} ~r~ z(t) \\$$
and its value is  is at most 0.2 pT: this excludes any chance of eddy currents, considering also the non-magnetic nature of Al5056.

The electromagnetic forces acting on the resonator are then:
$$F(t) = \frac{C_0 V_0^2}{2 d_{eq}}  - K_{em} z(t) + F_{I}(t) $$
   $K_{el}$, defined in sect.\ref{sec:normaltuning}, ranges 0.2-0.6 M~N/m, less than 10\% of the mechanical stiffness. 

%

We can also consider the radiated power:  we have a charge of $\sim 1-5 \mu$C oscillating at 1 kHz.  Using the Larmor formula we can evaluate:
 $$  P =  \mu_0 \frac{(\omega^2 z)^2}{ 6 \pi c} \sim 10^{-22} W    $$

Finally,  the e.m. pressure generated inside the gap $ p = \frac12  \varepsilon_0 E^2 $ has a time-dependent component, giving rise to an an a.c. force:
$$ F(t) =  S \cdot \varepsilon_0 E_0^2 \frac  {C_s}{C_0 + C_s} \frac{z(t)} {d_{eq}}  $$
This is obviously an alternative way to deduce the electrical stiffness $K_{el}$ of eq.\ref{eq:Kel}.

\subsection{Non-linear electric susceptibility of PTFE spacers}
\label{sec:ptfe}
A change in the stray capacitance $C_s$ could lead to an increased stiffness, as shown by eq.\ref{eq:tune}.
The main suspects are the PTFE spacers insulating the facing electrode:  the change in $C_s$ should appear as a consequence of a non-linear permittivity (voltage dependent dielectric constant).   Actually, PTFE thin sheet is a linear dielectric up to its breakdown field \cite{PTFE2}. Besides, we operate 
quite distant from the breakdown voltage of 80 kV/mm  \cite{PTFE}, reaching a maximum of 980 V/ 70$\mu m$ = 14 kV/mm  for dataset 4. 

\subsection{Residual gas in the gap}
\label{sec:gas}
As discussed above, the residual gas in the vacuum chamber is  $10^{-3} - 10^{-4}$ Pa ;  we can speculate that in the capacitor gap, due to reduced conductance, it might be ten times higher.  As the dielectric susceptibility of air is  $\varepsilon_r - 1 =  5.9 ~10^{-4} $ at NPT  \cite{aria} and it  scales roughly linearly with pressure, no significant change in the active capacitance $C$ can be expected.

\subsection{Uehling potential}
The Uehling potential is the lowest order correction to Coulomb field between two point charges, due to vacuum polarization caused by very strong electric fields at atomic distances.  Such corrections are order $\alpha_{fs}$, fine structure constant, with respect to the leading Coulomb potential \cite{Uehling}.
But at long-range such corrections should be of many orders less.

\subsection{Influence of electric field on the  stiffness of cantilevers}
\label{sec:stif}
One possibility we considered is a weakening in the elastic constant of the cantilevers when their surfaces are exposed to an electric field, even if we do not know of any mechanism that can produce this effect.
 Anyways, as shown in fig.\ref{fig:schema}, the cantilever arms are positioned at a much larger distance from the charged plate than the resonator disc:  we are dealing with mm, rather than $\mu m$ and the electric field strength is thus attenuated by a factor $\sim 10^3$. Besides, only a fraction of the cantilever is exposed to the electric field, the outermost part actually faces the grounded external disc. \\
 Electrostriction is an effect mostly visible in dielectrics, but it can also be observed on conductors: we evaluate its possible influence on the cantilevers and on the electrode.
 It is well known that an electrostatic field does not penetrate in a conductor but influences on its surface producing electrostriction effect: both stretching out (increase of volume) and deformation of the conductor. In our case the strain of resonator and portions of cantilevers is small (as shown below), so that only the effect of extending of their volumes  (without shape changing) needs to be considered. 
 As discussed in \cite{LL7b},  the fractional change in the conductor volume $\Omega_0$ can be related to material properties as:
 $$  \frac{\Delta \Omega}{\Omega_0} =  \frac{\Delta P} K   $$
 
 where $K$ is the bulk modulus of the material (Al 5056),  $\Delta P = \frac 12 \varepsilon_0 E^2 $ is the  stress energy on the metal surface of resonator or cantilevers. The increase of volume $\Delta \Omega = \Sigma \cdot \Delta h$ is due to change in the thickness of the conductors.
 In the electrostatic case, with $V \simeq 10^3$ V, we obtain $\Delta h \simeq 2$ nm for the vibrating electrode and $\Delta h \simeq 70$ fm for the cantilevers.  For a field oscillating at 1 kHz, the effect is much lower (see also \cite{KKZ}).  We conclude that the influence of electric field on the mechanical properties of  resonator and  cantilevers is virtually null.
 

 \subsection{Anharmonic and damping effects of oscillator}  
 \label{sec:nonlinear}
 In all  measurements, the resonant frequency of the transducer did not depend on the amplitude of excitation, that varied in a range of at least two orders of magnitude. The frequency was unchanged within the measuring accuracy dictated by the decay times of the resonator,  $\pm ~0.1$ Hz.
 This indicates that no anharmonic effects  are detectable in our tests.  
 
 Non-linear effects, i.e. contribution  from  second and higher order terms of eq.\ref{eq:force2} and depending on the vibration amplitude A,  can be estimated via the Poincar\`e- Lindstedt theory \cite{LL1}:
\begin{align*}
  f= &f_0 - \frac 5 {12\cdot2\pi}  \frac1{(2\pi f_0)^3} \frac1{(2 m)^2}
 \bigg[\frac{d^3 \mathcal E_{em}}{dz^3} \bigg]^2 A^2 =\\
=& f_0 - \frac{15}{256 \pi^4 f_0^3} \bigg[\frac{V^2}{m d_{eq}^3} \frac {C~C_s^2}{(C+C_s)^2}\bigg]^2  A^2
\end{align*}
Even assuming an unreasonably large vibration $A = 1 \mu$m, the frequency correction would amount to few $\mu $Hz in the various datasets.

Damping reduces the resonant frequency as $  f = f_0 \sqrt{1 - \frac 1{(2 \pi f_0 \tau)^2}}$, where $\tau$ is the decay time. It is also well known that $\tau$ decreases with the applied field, so that the change in frequency is enhanced at high fields. In our cryogenic runs, the decay time changed from 700 down to 50 s, while at room temperature the effect was more evident, reducing $\tau$ from 7 to 0.1 s.
Even in the worst case, $\tau = 0.1$ s,  the change of resonant frequency  is  $f- f_0 \sim -1~ $mHz, while the unaccounted shift is about -59 Hz.

\section*{Appendix C - Mechanical impedance}
\begin{figure}[htbp]
\begin{center}
\includegraphics[width=0.6\larghez]{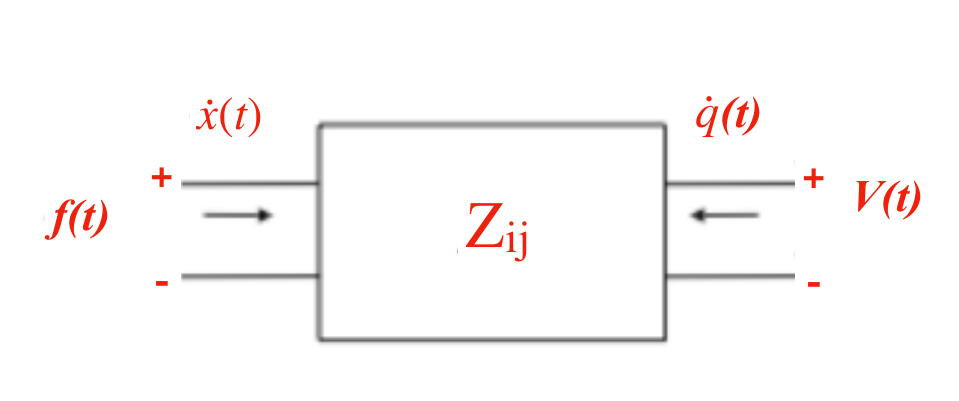}
\caption{Schematics of a linear transducer as a two-port}
\label{fig:twoports}
\end{center}
\end{figure}
An electromechanical transducer is often described in terms of a linear two-port, as shown in fig. \ref{fig:twoports}, by two equations relating mechanical variables (velocity $\dot x(t)$ and force $F(t)$) and electrical variables (current  $\dot q(t)$ and voltage $v(t)$). 

\begin{eqnarray*}
F &=  Z_{11} \dot x + Z_{12} \dot q \\
v & = Z_{21} \dot x + Z_{22}  \dot q
\label{eq:twoport}
\end{eqnarray*}
where the $Z_{ij}$ are linear (or linearized) differential operators, depending on the specific device.  $Z_{11}$ is the mechanical impedance (force/velocity), $Z_{22}$ the electrical input impedance, while $Z_{12}$ and $Z_{21}$ are the  reverse (force/current) and direct (voltage/velocity) transimpedence, i.e. the transducting terms. In linear devices we have $Z_{12}  = -Z_{21}^*$.
Usually, force and voltage are considered external variables (forcing terms) while velocity and current are internal.  By setting $v(t)=0$ (no time dependent electrical excitation) and eliminating $ \dot q(t)$ between the two equations, we obtain:
\begin{equation*}
F(t) = [ Z_{11} - \frac{ Z_{12}Z_{21}}{Z_{22}} ] \dot x(t)
\end{equation*}
The term in parenthesis is the modified mechanical impedance.  With $Z_{11} =i m \omega_0$ ($ i\equiv \sqrt{-1}$) and the values of matrix elements for a capacitive transducer:  $ Z_{12}= Z_{21}  = E/ i \omega $ and  $ Z_{22} =  [1/i \omega C + 1/i \omega C_s] $
we obtain,
 on resonance:
$$ Z_{11} (E) = i  m ~\omega_0  \bigg[1 - \frac {E^2}{m \omega^2_0} \frac {C C_s}{(C+ C_s) }\bigg]$$
that is the same relation as in eq. \ref{eq:tune}.

\section*{Acknowledgments}
 We are grateful to  Boris Altshuler, Valentin Bessergenev, Lawrence H. Ford, Steve Lamoreaux, Peter Milonni,  Roberto Onfrio, Giuseppe Pucacco, Luigi Rosa, Boris Shapiro, Renata Sisto, Andrei Varlamov,  Gregory Vereshchagin, David Vitali  for useful suggestions and stimulating discussions.

 \end{document}